\documentclass{emulateapj}

\usepackage{graphicx}
\usepackage{gensymb}
\usepackage{epsfig}
\usepackage{epstopdf}
\usepackage{times}
\usepackage[]{natbib}
\usepackage{url}
\usepackage{color}
\usepackage{amssymb,amsmath}

\shorttitle{MHD avalanche heating}
\shortauthors{Hood et al.}

\begin{document}


\title{An MHD avalanche in a multi-threaded coronal loop}

\author{A. W. Hood$^{1}$, P. J. Cargill$^{1,2}$,  P. K. Browning$^{3}$ and K. V. Tam$^{1,4}$}
\affil{$^1$ School of Mathematics and Statistics, University of St. Andrews, St. Andrews, Fife, KY16 9SS, U.K}
\email{awh@st-andrews.ac.uk}
\affil{$^2$ Space \& Atmospheric Physics Group, The Blackett Laboratory, Imperial College London, Prince Consort Road, London, SW7 2AZ, U.K.}
\affil{$^3$ School of Physics and Astronomy, University of Manchester, Oxford Road, Manchester, M13 9PL, U.K. }
\affil{$^4$ Space Science Institute, Macau University of Science and Technology, Avenida Wai Long, Taipa, Macau}
 
\begin{abstract}
For the first time, we demonstrate how an MHD avalanche might occur in a multi-threaded coronal loop. Considering
23 non-potential magnetic threads within a loop, we use 3D MHD simulations to show that only one thread needs to be unstable in order to start an avalanche 
{even when the others are below marginal stability}.
{This has significant implications for coronal heating in that it provides for energy dissipation with a trigger mechanism.} The instability of the unstable thread follows the evolution determined in many earlier investigations. However, once one stable thread is disrupted, it coalesces with a neighbouring
thread and this process disrupts other nearby threads. Coalescence with these disrupted threads then occurs leading to the disruption of yet more threads as the avalanche develops.
Magnetic energy is released in discrete bursts as the surrounding stable threads are disrupted. The volume integrated heating, as a function
of time, shows short spikes suggesting that the temporal form of the heating is more like that of \textit{nanoflares} than of constant heating. 
\end{abstract}



\section{Introduction}
 The physical mechanism responsible for energy release in the solar corona remains unknown, 
but must involve the rapid dissipation of magnetic energy that is either stored in situ, or injected as
waves. In the former case, it is becoming widely accepted that the energy release does not
occur at a single site (e.g. a monolithic current sheet), but involves dissipation over
a large volume, a consideration motivated originally by the particle acceleration requirements
in solar flares. However, in view of the power law distribution of event sizes over a 
broad range of energies, such considerations can also be assumed to apply to smaller events
such as microflares and hypothesised nanoflares.

One example of such large-scale energy release of importance to this paper are the recent studies
of the magnetohydrodynamic (MHD) kink instability \citep{browning2008, hood2009, bareford2013} in a single twisted 
magnetic flux strand. Hood et al. (2009) showed that the magnetic energy is released as one 
extended \lq event\rq , which may correspond to a microflare or swarm of nanoflares.
While the initial stages correspond to the fast release of energy when a helical current sheet 
forms, the sheet then fragments and the plasma appears turbulent with many small current sheets
forming throughout the loop cross section. These sheets also release their energy as the magnetic 
field relaxes towards the final lowest energy (Taylor) state with a constant alpha force-free field.
Temperatures up to around $10^{7}$K result from Ohmic and viscous
heating \citep{tamphd}. {The importance of this result is that it provides a triggering mechanism for the release of
magnetic energy. Prior to the onset of the ideal MHD instability, the magnetic field is in a stressed but stable configuration. Current sheet
formation occurs during the non linear development.}

During the relaxation process, the loops expands in the radial direction until it is approximately 
1.5 times its original value. Hence, we expect that an unstable loop could influence
any nearby neighbours if they are within this new expanded radius. \citet{tam2015} investigated the conditions 
under which a nearby stable loop could be disrupted by an unstable loop. They showed
that the disruption could occur when the second loop was sufficiently close. The addition of other closely spaced loops sets up the intriguing possibility 
of initiating an MHD avalanche.

{\cite{lu1991} proposed that the solar coronal magnetic field is in a state of Self Organised Criticality (SOC), analogous to a sandpile onto which sand is being slowly poured, 
which naturally produces a power law distribution for the dependence of   solar flare magnitude  on occurrence frequency.   In the SOC paradigm, a local instability at a site occurs if a 
critical parameter value is exceeded: the subsequent reconfiguration releases some energy and also affects neighbouring sites, possibly causing critical conditions to arise there, with   
an ÒavalancheÓ resulting as the disturbance spreads out.  In the standard approach, the system is subject to random external driving during the intervals between avalanches, with 
SOC being reached when a statistical balance between external energy input and the dissipation is achieved.   Models are based on simple ÒrulesÓ  to determine when instability is 
reached and how the field subsequently reconfigures, usually with a cellular automaton (CA) approach.}
 
 {This approach to modelling solar flares, and solar coronal heating through nanoflares,  has been developed extensively   (see reviews \cite{charbonneau2001} and \cite{aschwanden2014}). 
 Different rules may be proposed for driving, for determining instability, and for relaxation to a stable state.  Typically, models  use a CA approach,  with  the magnetic field  represented 
 by  vector or scalar values on a 2D or 3D rectangular grid; relaxation is  triggered if a field value differs too much from  neighbouring values   (representing reconnection onset  
 above a critical field gradient) or if the horizontal field or twist exceed  a critical value  \citep{lu1991,lu1993,vlahos2004,lopezfuentes2010,lopezfuentes2015}.  However, \cite{hughes2003} consider an ensemble of 
 semi-circular field lines, giving a discrete model not using a rectangular lattice.  \cite{morales2008} and \cite{morales2009}  present a CA approach based on magnetic field lines rather than magnetic field 
 values on a grid,  guaranteeing  a divergence-free magnetic field.  
 \cite{strugarek2014} develop avalanche models with deterministic driving representing slow twisting of a loop, with the stochasticity required for SOC introduced through 
 the criticality or relaxation algorithms.   A CA model in which the instability criterion is inspired by kink instability of twisted loops has recently been proposed by \cite{mendoza2014}; 
 however, this model does not consider the reconnection between neighbouring loops,  which is predicted by our model.}
 
{SOC models have the advantage of being simple, with readily understandable physical principles and  the capability of  modelling  large volumes and time periods efficiently. 
 They generally  predict power-law distributions of event sizes, consistent with observations of solar flares. However, the assumed rules for instability onset and relaxation  
 are not derived directly from the underlying MHD equations. For example, it is difficult directly to relate the relaxed state in CA models to the predictions of MHD models \citep{taylor1974,yeates2014}.}
 
{In this paper, we predict an avalanche-type energy release using an ab initio MHD approach. It is not our aim to incorporate ongoing driving, which would  be computationally 
 unfeasible at this time.  We,  thus,  consider an initially stressed field with only a single avalanche, in contrast to the discrete models discussed above,  which are repeatedly 
 driven through a sequence of avalanches towards  SOC state.} While earlier MHD models are restricted to 2D and use the approximate equations of Reduced MHD 
\citep{dmitruk1998,georgoulis1998}, we present fully 3D results that place no artificial restrictions on the 
evolution of the magnetic field.

\begin{figure}[htbp]
\begin{center}
\includegraphics[width=0.4\textwidth]{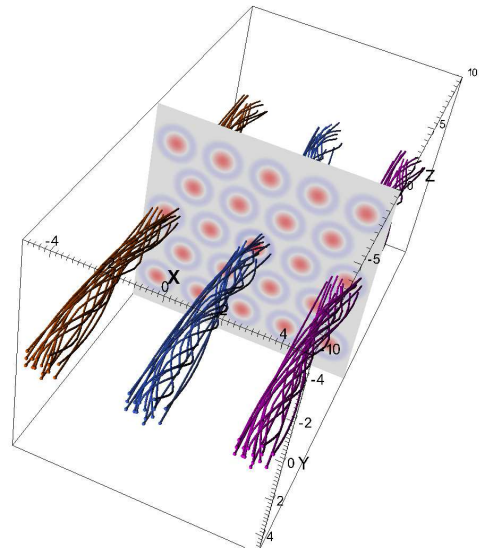}
\caption{{Twenty three threads are used in the avalanche simulation. The twisted field lines outline just three of the threads. The contours of the axial current density are shown in the mid-plane. 
The gaps between the threads are filled with a uniform axial field.}}
\label{fig:initialstateB}
\end{center}
\end{figure}

To the present date, there has never been a demonstration of how an avalanche can arise from first principles
using the full equations of 3D MHD. It is the purpose of this paper \textit{to demonstrate the proof of principle of an MHD avalanche}. We build on the work of \citet{tam2015} and 
investigate how a single coronal structure (e.g. a loop or part of an AR) that consists of 23 individual magnetic threads evolves in time when there is initially only one unstable thread 
surrounded by 22 non potential but stable threads. The basic model is described in Section 2 and illustrated in Figure~\ref{fig:initialstateB}. In Section 3 we demonstrate that an avalanche 
involving the majority of the threads can arise on rapid timescales. {Section 4 presents our conclusions.}

\section{Initial Setup}\label{sec:setup}
We assume that the magnetic field in a coronal loop consists of a number of threads. In our case we take 23 threads. Each thread is described below and is 
non potential. However, only one thread is actually unstable to the ideal MHD kink instability. The other threads are
well below the marginal stability threshold. {Thus, we initiate the MHD avalanche through a kink instability in a single thread. This is equivalent to the 
ideas of CA that reconfigure fields once a local gradient (or other quantity) exceeds a critical value.}

We solve the resistive MHD equations using the Lagrangian Remap code, Lare3D, as described in \citet{arber2001}. Shock viscosities are used to treat shocks and the Ohmic and shock heating
are added to the thermal energy equation \citep{bareford2015}. To keep the results simple, we ignore thermal conduction and optically thin radiation. The effects of these terms have been considered 
by \citet{botha2011}, \citet{bareford2015a} and \citet{tamphd}. There
are no additional ad hoc heating terms included, so that all the heating comes from the dissipation of energy in shocks and Ohmic dissipation. To avoid any additional diffusion of the background current, the
resistivity is zero and is only switched on if the magnitude of the current density exceeds a threshold value. This is discussed in, for example, \citet{hood2009}, where the non-dimensionalisation is also described. 
We select a length scale
of $a_0 = 1$ Mm, based on the radius of each thread, a field strength of $B_0 = 10$ Gauss and a density of $\rho_0 = 1.67 \times 10^{-12}$ kg\  m$^{-3}$ that corresponds to a number density of $10^{15}$m$^{-3}$.
The reference time is {1.45 seconds, based on} the Alfv\'en travel time across a thread.

The number of grid points is limited by the computational resources available and we used $480$ points in both directions of the loop cross section and $960$ along the loop length. The boundaries are located
at $x = \pm 5$, $y = \pm 5$ and $z = \pm 10$. All boundaries are taken as line-tied so that
there are no velocities on any of the boundaries. 

The single loop analysis has shown how the instability depends on numerical resolution and boundary conditions. \citet{tam2015} showed that insufficient resolution results
in some numerical diffusion of the current and an unstable loop is destabilised without any initial disturbance. In our case, each
thread has approximately 100 grid points across its diameter and this is sufficient to reduce the influence of numerical diffusion. The side boundaries and the location of the loop footpoints tend
to stabilise the kink instability if the sides are too close to the loop and the loop length is too short. However, what is clear is that any instability in these stabilising circumstances
means that there will be an instability when the boundary conditions are relaxed.

\subsection{Initial Equilibria}
Consider the coronal situation, where the ratio of the gas pressure to the magnetic pressure is so small (around $10^{-3}$), so that the magnetic field can be assumed to be force-free. 
Hence, $\nabla \times \textit{\textbf{B}} = \alpha \textit{\textbf{B}}$. 
For simplicity of modelling, we assume that each equilibrium magnetic thread can be modelled by a straight twisted cylinder, with the cylinder axis of the $i^{th}$ thread located at $(x_i, y_i)$, and we use the smooth $\alpha$ profile 
that only depends on radius, $r$, as described by \citet{hood2009}. Therefore, each magnetic thread is non potential but has zero net axial current, corresponding to localised twisting.
For a radial coordinate defined by $r^2 = (x-x_i)^2 + (y - y_i)^2$, {where the location of the axis of the thread is $(x_i, y_i)$,} the magnetic field components of the $i^{th}$ thread, for $r \le 1$,  has the form
\begin{eqnarray}
 B_{\theta} &=& B_i \lambda_{i} r (1-r^2)^3, \label{eq:b1}\\
 B_z &=& B_i \sqrt{1 -\frac{\lambda_i^2}{7} +\frac{\lambda_i^2}{7} (1-r^2)^7 -\lambda_i^2 r^2 (1-r^2)^6},  \label{eq:b2}\\
 \alpha &=& \frac{2 \lambda_i (1-r^2)^2 (1-4 r^2)}{B_z}, \nonumber
\end{eqnarray}
and, for $r > 1$, 
\begin{eqnarray}
 B_{\theta} &=& 0, \label{eq:bb1}\\
 B_z &=& B_i\sqrt{1 - \lambda_i^2}/7,  \label{eq:bb2}\\
 \alpha &=& 0. \nonumber
\end{eqnarray}
$B_i$ is the magnetic field strength on $r=0$ {of the $i^{th}$ thread} and $\lambda_i$ is a constant parameter related to the twist in the magnetic field. Thus, the space between the threads is filled with axial (untwisted) field; 
note that fields are continuous everywhere, and in force balance. The maximum value of $\lambda_i$ is restricted by the fact that 
$B_z^2$ must be positive and, therefore, $\lambda_i <  2.438$. $\lambda_i$ controls the stability properties of the thread and the marginal stability value, $\lambda_{crit}$,
does depend on the length, $2 L_{z}$, of the thread. For our case, $2 L_{z} =20$ and $\lambda_{crit} = 1.586$.  The stability threshold for longer threads will be given by smaller values of $\lambda_i$ \citep{bareford2010}. If the system can create an avalanche with short threads, there will definitely be an avalanche for longer, less stable threads.
$\lambda_i$ also controls the maximum value of the magnitude of the {initial} current, i.e. $2 \lambda_i B_i$. Note that $\lambda_{i}$ is positive in each thread so that they all
have the same sense of rotation.

\begin{figure}[htbp]
\begin{center}
\includegraphics[width=0.4\textwidth]{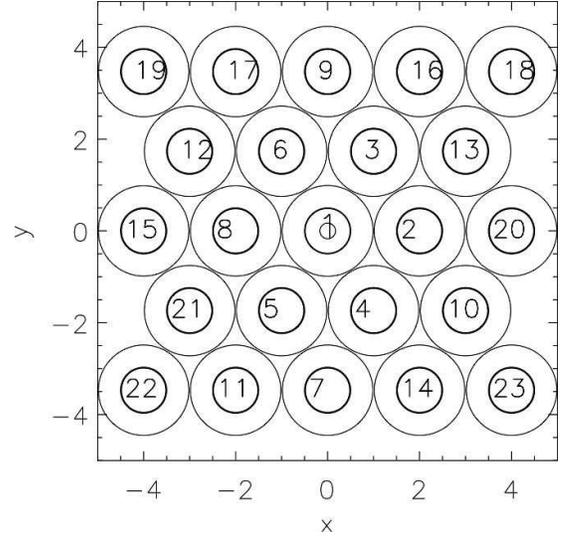}
\caption{The 23 threads are outlined by contours of axial current density and numbered for identification. Only loop $1$ is unstable. The numbers correspond to the order in which they are disrupted and this is
used in the avalanche description in the main text.}
\label{fig:initialstate}
\end{center}
\end{figure}
The critical current used in switching on the anomalous resistivity is always chosen to be larger than {the maximum value}. The axial flux within a thread is of the form $\Phi = 2\pi B_i f(\lambda_i)$,
where $f(\lambda)$ is a monotonically decreasing function of $\lambda$. {Thus, the magnitude of the magnetic field on the axis of each thread is $B_i$ and this} 
is proportional to the axial magnetic flux in the thread. Note that the values of $\lambda_i$ and $B_i$ are related 
for the different threads. From Equation (\ref{eq:bb2}), a smaller value of $\lambda_i$
requires a smaller value of $B_i$ so that all threads are embedded in the same uniform potential field. 

The initial array of magnetic threads is shown in Figure~\ref{fig:initialstate}. This has a similar but superficial appearance to the lattice used by \cite{charbonneau2001}, where each node was assigned
a value $B_{k}$ as a measure of the magnetic energy. However, it is important to remember that {in this paper} the magnetic field is continuous and that a small sample of contours for the
axial current density are chosen to aid in the identification of each thread {in Figure~\ref{fig:initialstate}}.
Contours of the axial current are shown at the mid plane $z=0$. The
individual threads are numbered by the order in which they are disrupted. The unstable thread is labelled as \lq 1\rq {and has $\lambda_1 = 1.8 > \lambda_{crit}$. All the remaining threads
have the same value of $\lambda_{i}$ given by $\lambda_{i} = 1.4 < \lambda_{crit}$. Thus, all the stable threads have a $\lambda$ value that is significantly below the marginal value {of $\lambda_{crit} \approx 1.6$.}
{In order to emphasise that this simulation is using a continuous magnetic field, the initial axial field, $B_{z}$, and the initial $B_y$ component are shown in Figure~\ref{fig:Bzinitial} as functions of $x$, 
at the mid plane, $z=0$, and at $y=0$. Five
threads in this cut are clearly visible and the larger values of $B_z$ and $B_y$ indicate the unstable thread.}

Since the radial profile of $\alpha$ has both positive and negative values, the total magnetic helicity in the equilibrium field will be relatively small. In this case, the Taylor relaxed state will be weakly twisted and close
a potential field. Hence, in the mid-plane $z=0$, the magnetic fields will eventually evolve to a 
nearly uniform field in the axial direction. $B_x$ and $B_y$ are small and will only be significantly larger near the photospheric boundaries, $z=\pm L_z$.
\begin{figure}[htbp]
\begin{center}
\includegraphics[width=0.5\textwidth]{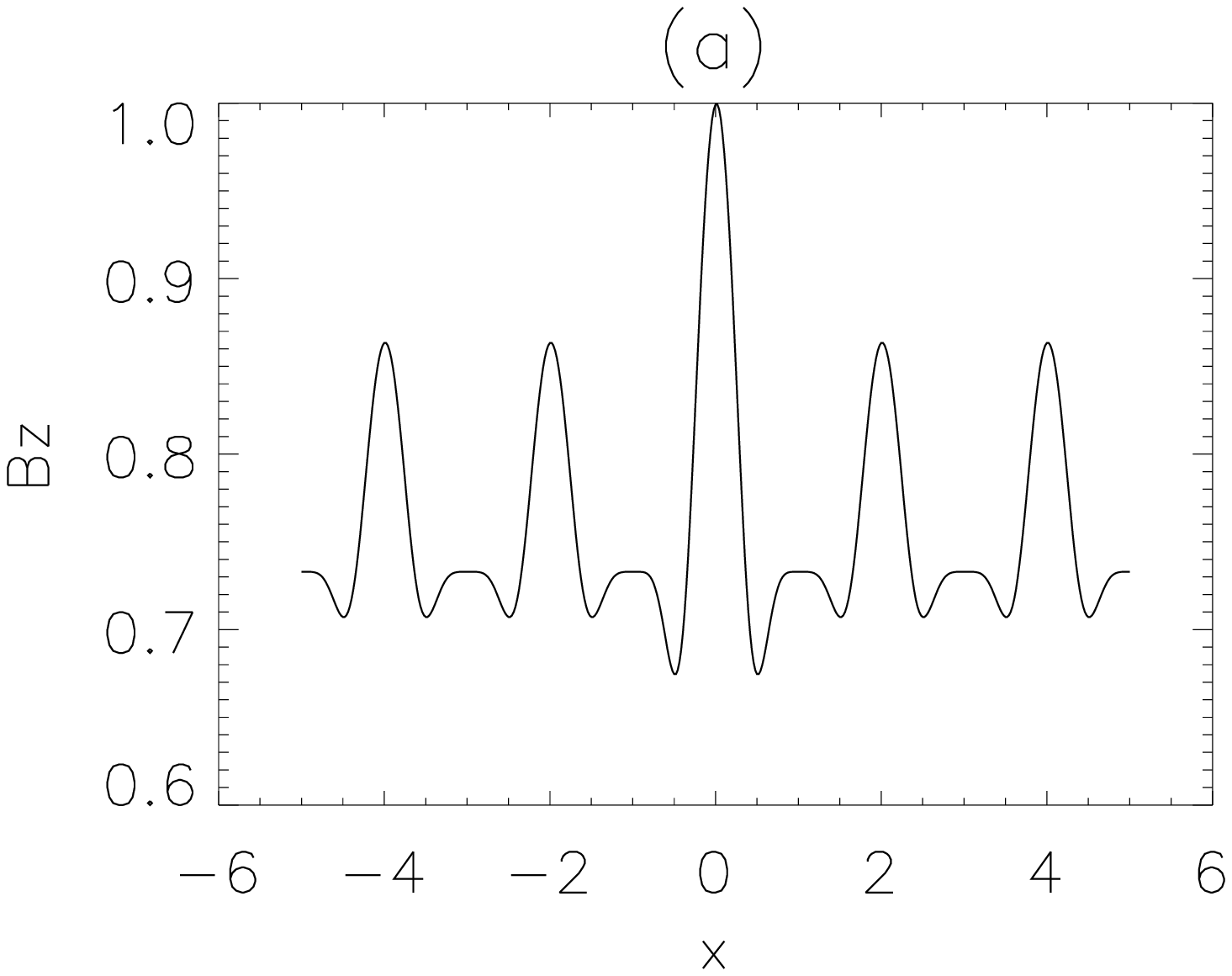}
\includegraphics[width=0.5\textwidth]{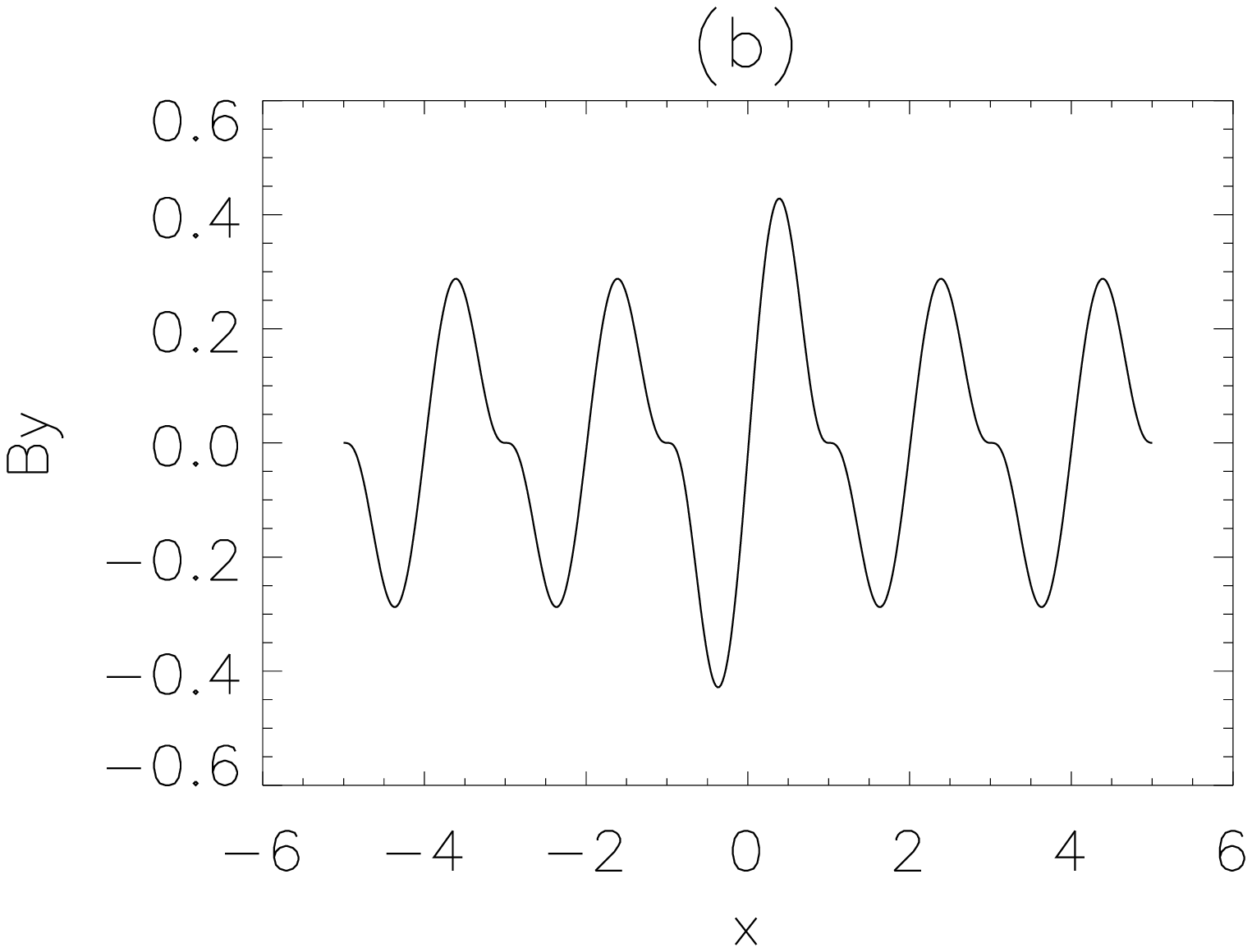}
\caption{The initial (a) axial field, $B_z$, and (b) transverse field,$B_y$, at the mid plane ($z=0$) and at $y=0$ as a function of $x$.}
\label{fig:Bzinitial}
\end{center}
\end{figure}

\section{Results}\label{sec:results}
To initiate the simulation, a small velocity disturbance is applied to the unstable thread, as discussed in \cite{tam2015}. Note that if this disturbance is
applied to a stable thread nothing happens. Hence, all except one thread are stable to small disturbances. {Remember that the magnetic field varies continuously across each thread and that each thread is surrounded
by a potential field with a constant axial field component.}
\subsection{Axial Current Density}
\begin{figure}[htbp]
\begin{center}
\includegraphics[width=0.4\textwidth]{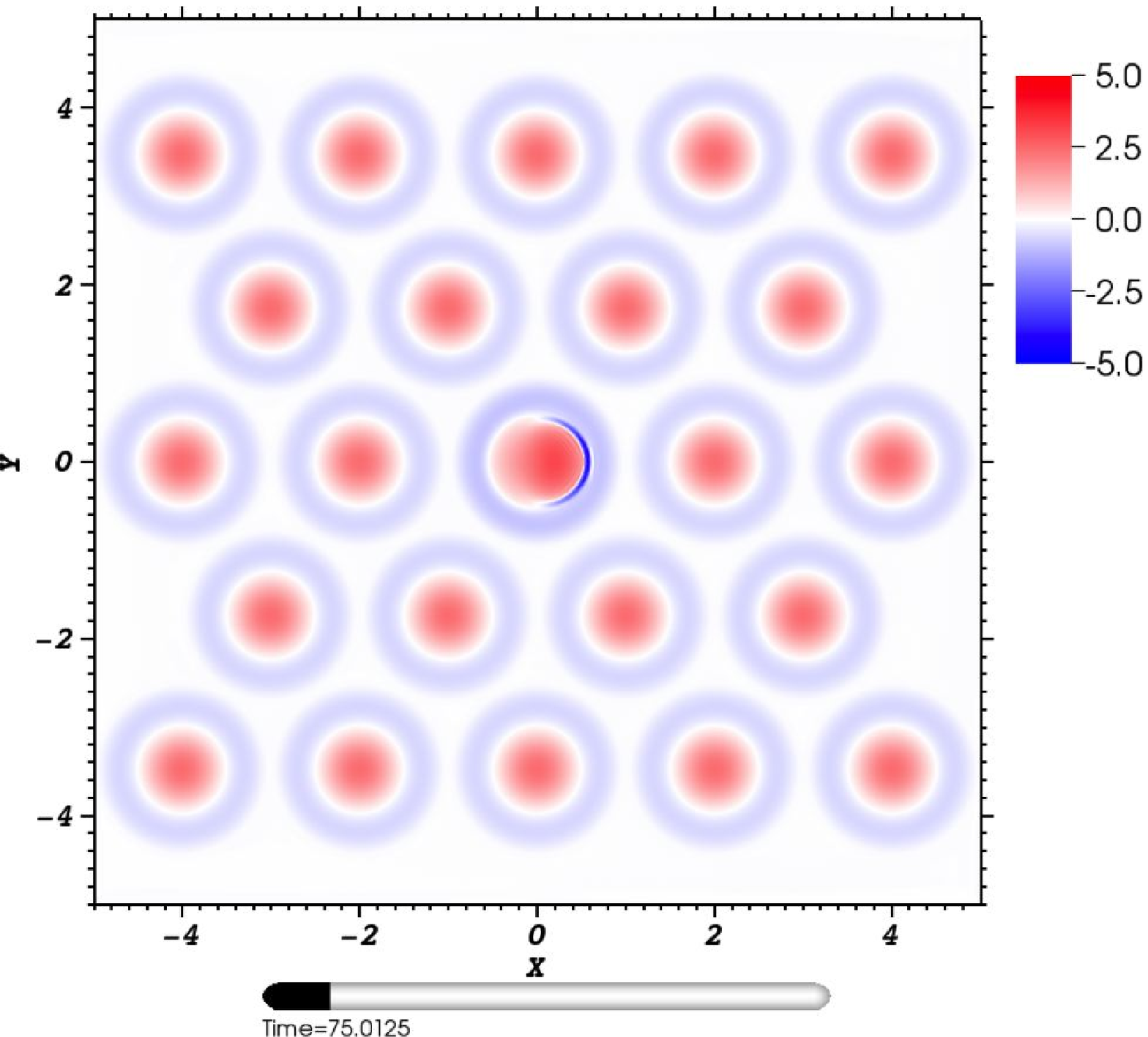}
\includegraphics[width=0.4\textwidth]{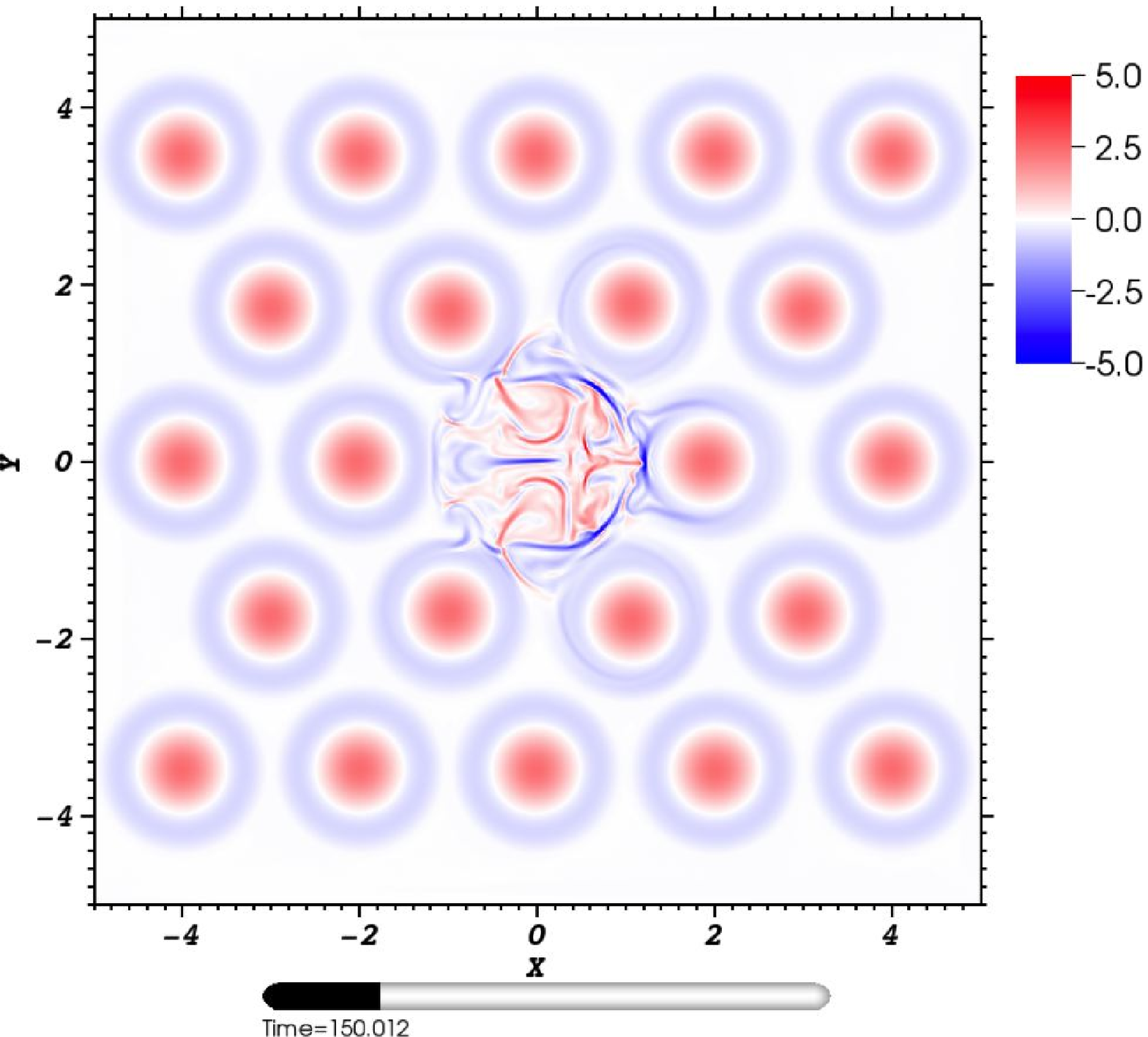}
\includegraphics[width=0.4\textwidth]{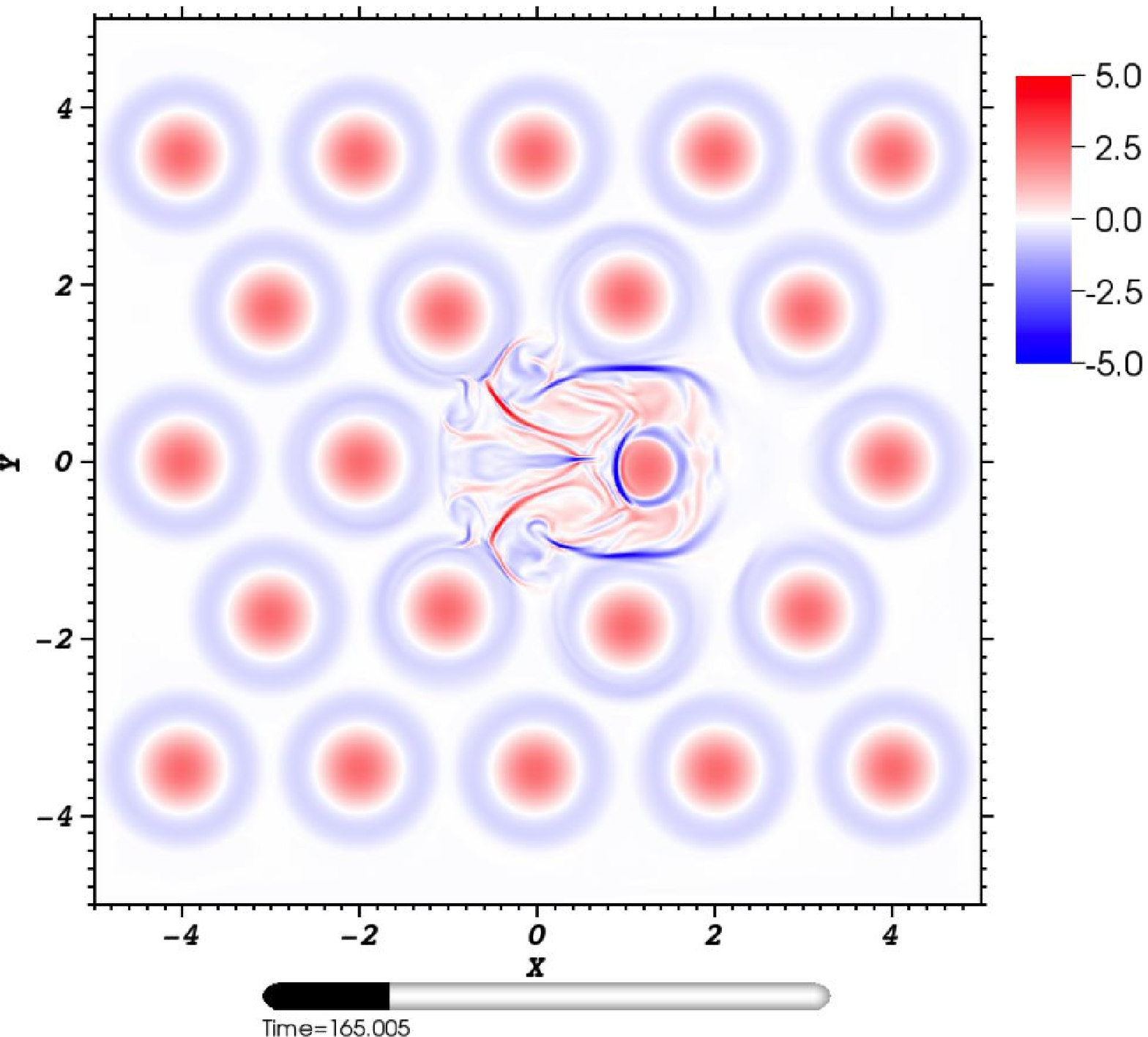}
\caption{Contours of the axial current at the mid plane ($z=0$) for $t=$ 75, 150, 165 during the early stages: stage 1 (top and middle panels) and stage 2 (bottom panel). Here the background resistivity is zero. 
Red corresponds to positive current, blue to negative and white to zero.}
\label{fig:jzmanymedres1}
\end{center}
\end{figure}
\begin{figure}[htbp]
\begin{center}
\includegraphics[width=0.4\textwidth]{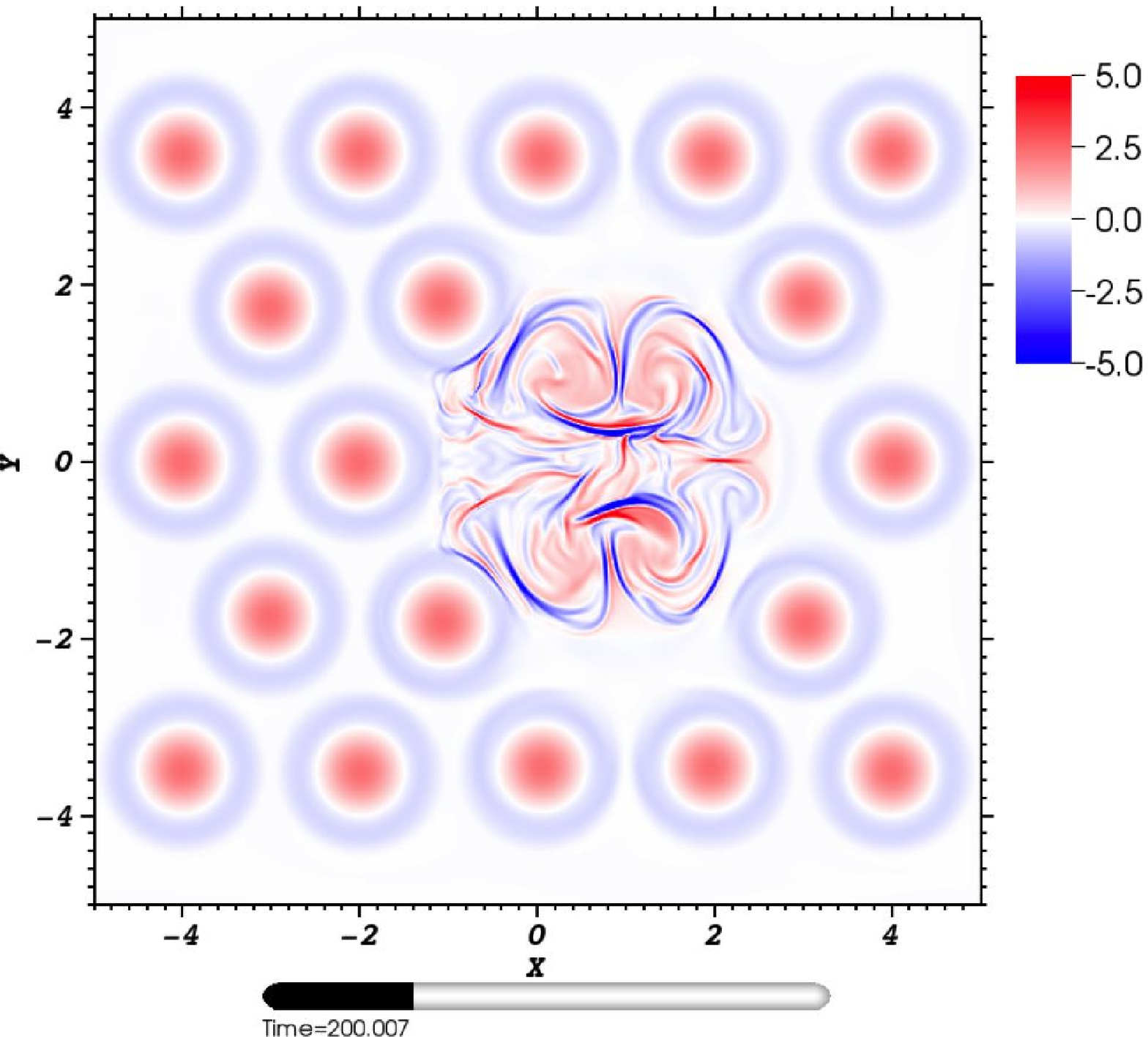}
\includegraphics[width=0.4\textwidth]{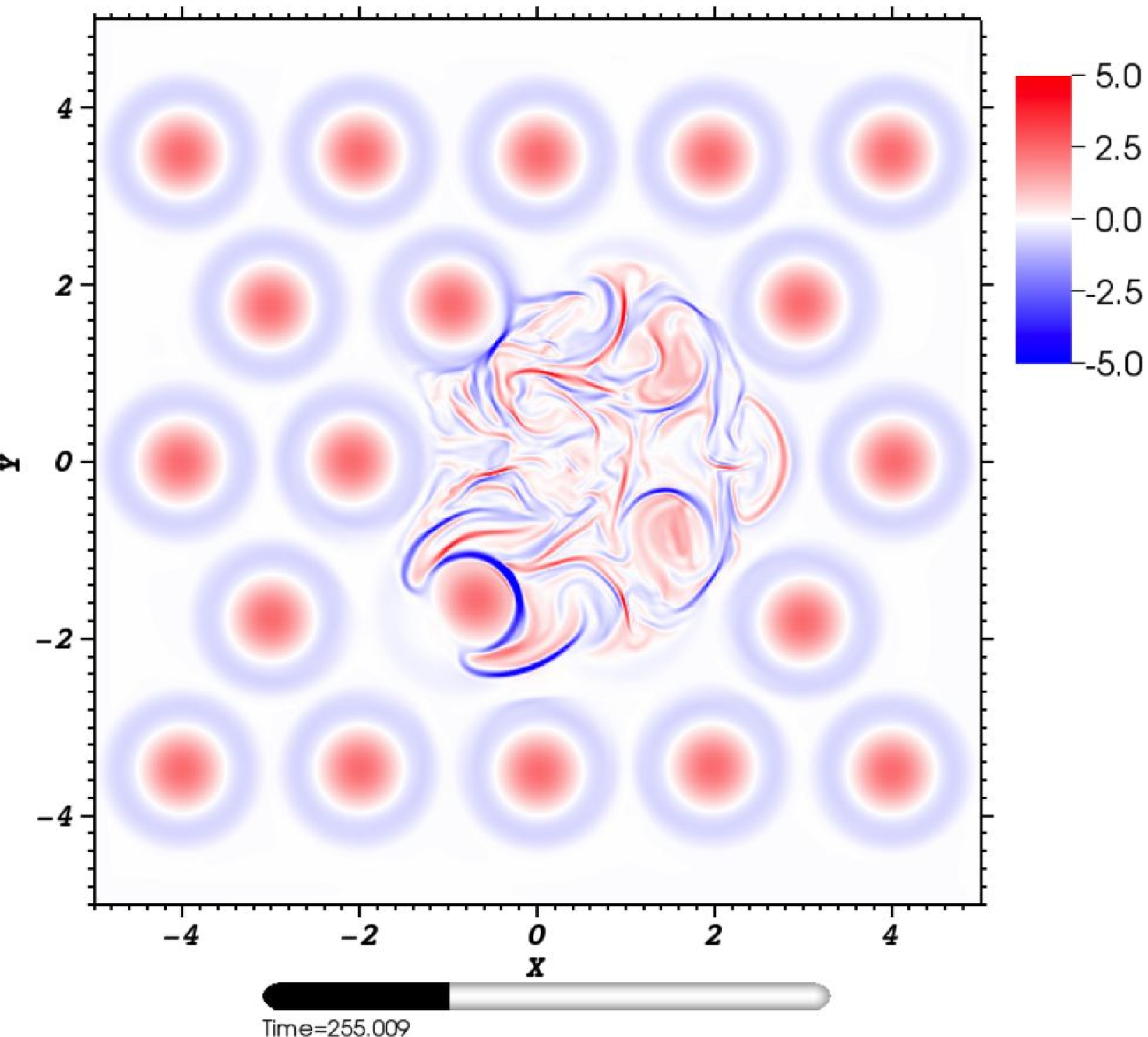}
\includegraphics[width=0.4\textwidth]{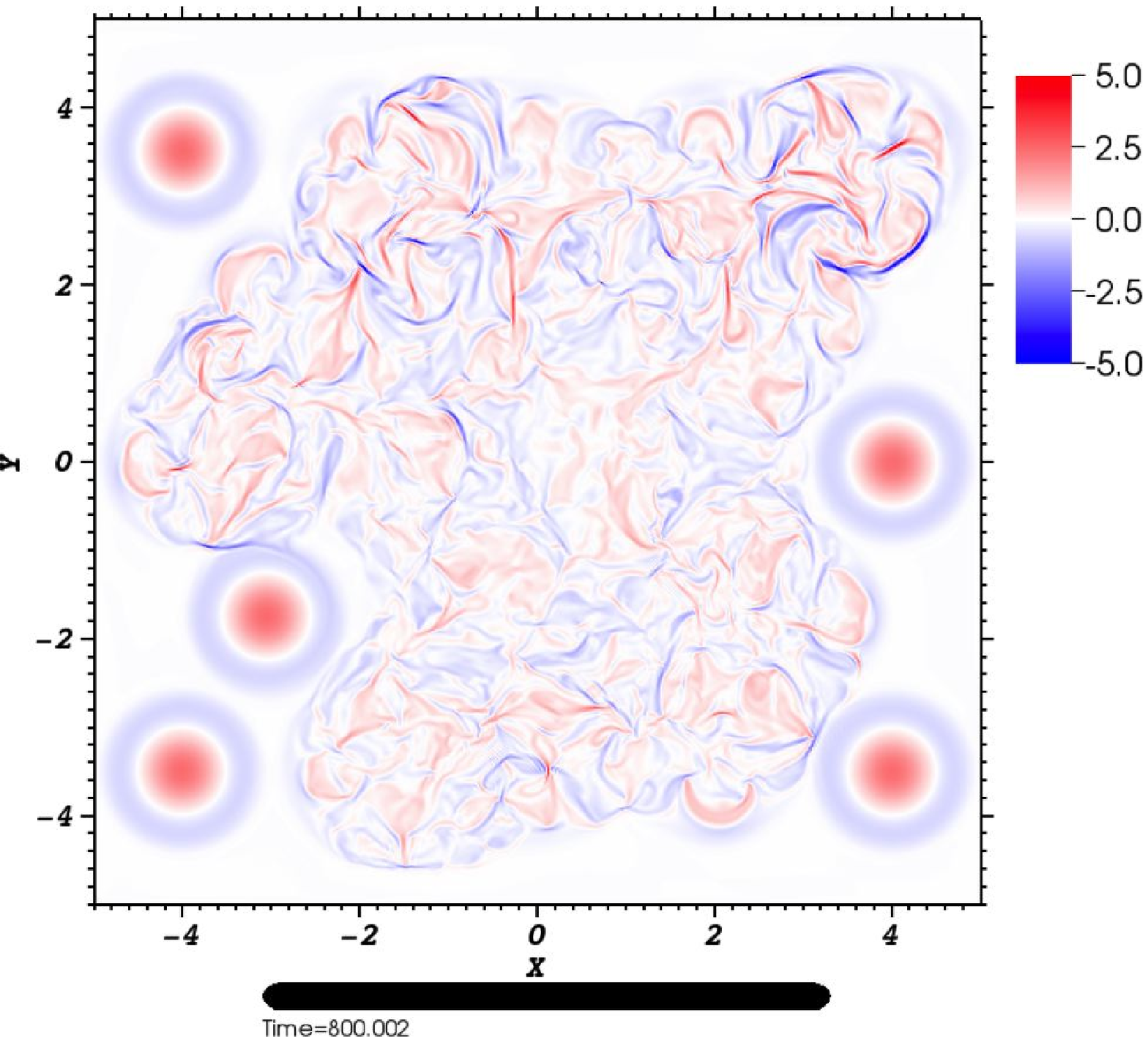}
\caption{Contours of the axial current at the mid plane ($z=0$) for times 200, 290, 800 during the later stages. Stage 3 is shown in the top panel, stage 4 in the middle and stage 9 in the bottom. 
Here the background resistivity is zero. Red corresponds to positive current, blue to negative and white to zero.}
\label{fig:jzmanymedres2}
\end{center}
\end{figure}
To illustrate the development of an MHD avalanche, we show results in the cross section at the mid-plane, $z=0$. Contours of the axial current, $j_z$, 
give a clear indication of current sheet formation, break up and relaxation. The simulation runs to a final time of $t=800$. During this period, there are 9 distinct interaction stages 
that lead to the MHD avalanche.
These are shown in Figures~\ref{fig:jzmanymedres1} and {\ref{fig:jzmanymedres2}. Figure~\ref{fig:jzmanymedres1}
shows contours of $j_z$ at three early times and covers stage 1 (linear phase, top), stage 1 (non-linear phase, middle) and stage 2 (first interaction). 
The top panel is during the initial kink instability of loop1 at time $t=75$. The first current sheet is clearly seen in loop 1. The middle panel is at time $t=150$. The
current sheet in the unstable loop has fragmented and small current sheets are forming throughout the volume of loop 1. This is during the first stage of the avalanche, as the single unstable thread is evolving. The 
fragmented currents are reminiscent
of a turbulent system, with a range of scales all the way down to dissipation.
In addition, thread 1 has expanded and is now interacting with the neighbouring threads, numbered
2, 3, 4, 5, 6 and 8 as indicated in Figure~\ref{fig:initialstate}. However, the first stable loop to be disrupted is thread 2, as can be seen in the bottom panel in Figure~\ref{fig:jzmanymedres1} and this is the start of stage
2. Because of the fragmented current sheet
development of thread 1, it is not clear why it should be thread 2 that is disrupted first, although this loop is closest to the initial midplane current sheet in thread 1 
(see top panel of figure 2). Note that as thread 2 begins to coalesce with thread 1, it interacts strongly with threads 3 and 4, forming current sheets between
them. At the mid-plane the $j_z$ contours are still remarkably symmetric, despite the kink instability breaking symmetry and the turbulent nature of the plasma.

Interactions then start to happen more frequently, as shown in Figure~\ref{fig:jzmanymedres2}. The top panel shows $j_z$ contours at time $t=200$. This is stage 3 and threads 3 and 4 have been pulled into the middle 
and there is evidence of current sheets forming at
threads 5 and 6. The symmetry in the interactions is no longer seen at time {$t=290$ (middle panel)} and thread 5 is already beginning to disrupt before thread 6 in stage 4. The following stages are noted. In stage
5, thread 7 is disrupted, essentially on its own. Stage 6  involves threads 8, 9 and 10 (in that order). Stage 7 sees the disruption of threads 11, 12, 13 and 14. Stage 8 covers the disruption of threads 15 and 16 and the final stage
in this simulation {(bottom panel)} involves the disruption of threads 17 and 18.

There may be a few more interactions after the end of the simulation but it is also possible that the avalanche ends with a few threads still remaining unaffected. In this simulation those unaltered are threads 19, 20, 21, 22 
and 23.

\subsection{Magnetic field lines}
Figure~\ref{fig:blines} shows a sample of the magnetic field lines for three
times: namely the initial state, the partially relaxed state at $t =
400$ and the state at the end of our simulation ($t = 800$). The
individual twisted threads are clearly seen in the top figure
and as the threads are disrupted the twist is reduced, leaving
only weakly twisted field lines at the end. The system has not
yet fully relaxed, {as there are 5 threads that are still twisted and in their initial state.}
\begin{figure}[htbp]
\begin{center}
\includegraphics[width=0.4\textwidth]{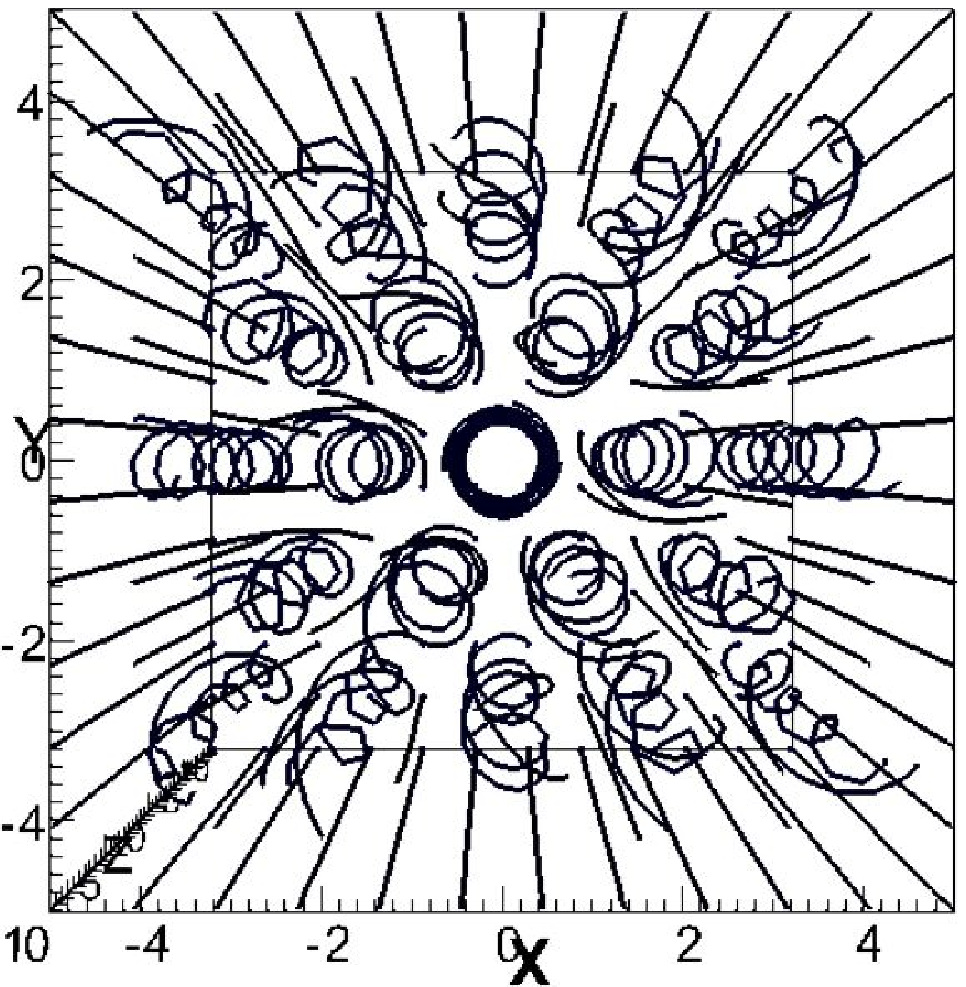}
\includegraphics[width=0.4\textwidth]{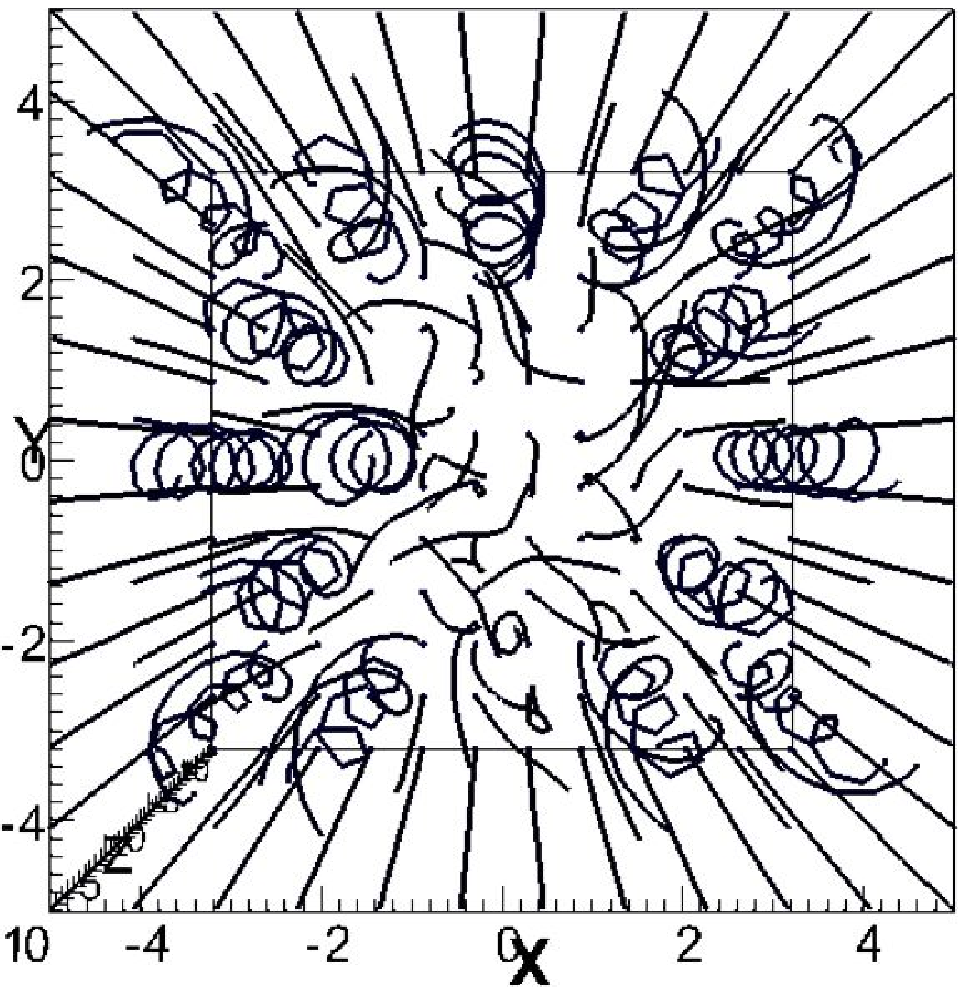}
\includegraphics[width=0.4\textwidth]{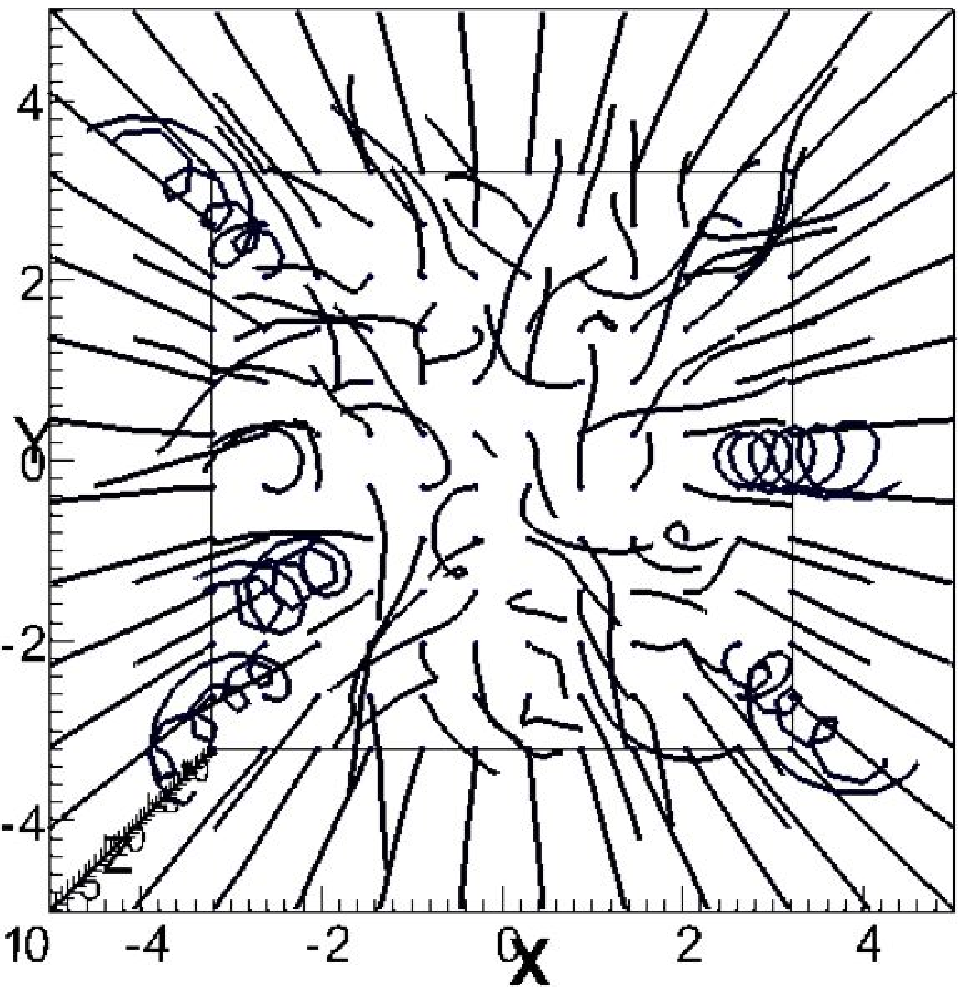}
\caption{{Magnetic field lines shown for times 0 (top), 400 (middle) and 800 (bottom) showing the initial state, halfway through the simulation and the final time.}}
\label{fig:blines}
\end{center}
\end{figure}

\subsection{Heating}
As one would expect, the magnetic energy is reduced at each of the stages when a thread is disrupted. The volume integrated magnetic energy, minus its initial value, is shown as a function of 
time in Figure~\ref{fig:totalenergy}. There is no reduction in magnetic energy until the kink instability develops around $t=75$. Then, there are several times when the gradient is steep and then followed by a shallower
gradient. These periods of rapid decrease in magnetic energy correspond to the various loops being disrupted.

The maximum free magnetic energy of each thread depends on the value of the twist parameter, $\lambda$, and this can be estimated as its initial energy minus the potential field with the same axial flux. Strictly
speaking the radius of the relaxed thread will be larger than the original thread but we ignore this small effect. For the unstable thread, $\lambda = 1.8$ and the free volume integrated magnetic energy is 1.8. 
From Figure~\ref{fig:totalenergy} the magnetic energy released is approximately 1.3 between $t=75$ and $t=145$. Not all the available energy is released in this time partly because the relaxed state is not a potential
field and partly because the relaxation process takes longer to reach its final equilibrium. Stage 2 is triggered before this can happen. For the stable threads the volume integrated magnetic energy available is 0.8
and during stage 2, the magnetic energy of the thread is reduced by 0.6. Again not all the energy is released. Similar amounts of energy are released during the other stages. For example, stage 7
lasts  between $t=490$ and $t=600$ and
the magnetic energy is reduced by 2.8 units. There are 4 threads disrupted during this period so that around 0.7 units of magnetic energy is released per thread.

\begin{figure}[htbp]
\begin{center}
\includegraphics[width=0.45\textwidth]{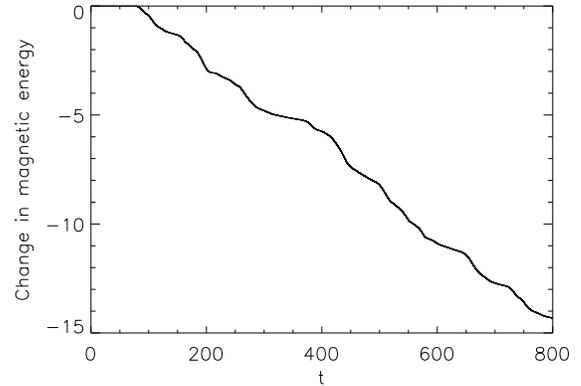}
\caption{The change in the volume integrated magnetic energy as a function of time. Each rapid change in energy corresponds to a significant release of magnetic energy. The dashed vertical lines indicate the 
start of the different stages, as determined from the $j_z$ contour plots.}
\label{fig:totalenergy}
\end{center}
\end{figure}

The time derivative of the volume integrated magnetic energy is related to the Ohmic heating. However, it is better to calculate
the volume integrated heating due to both Ohmic and viscous heating, where the viscous heating is due to shock heating, which 
has shown to be a very important effect \citep{bareford2015}. The volume integrated heating function, $H(t)$, is plotted in Figure~\ref{fig:heat}
as a function of time. This is the form of heating function that should be used in the single field line modelling of coronal loops. From this
figure, the 9 different stages can be identified. Stage 1 starts around $t=75$. The start
times of each of the stages can be estimated from the $j_z$ contours. The start of a stage is determined by the time that the first thread in a group starts to move. 
The start times are plotted as vertical dashed lines in Figure~\ref{fig:heat}. They are always just when the heating function starts to rise rapidly.

The heating function is clearly not uniform but instead consists of bursts of heating with the amplitude of the burst depending on the number of threads that are disrupted around the same time. In addition, the length
of time of the burst of heating also depends on the number of threads involved.

Note that there is a low background level of Ohmic heating during the simulation, after the onset of the initial instability.

\begin{figure}[htbp]
\begin{center}
\includegraphics[width=0.45\textwidth]{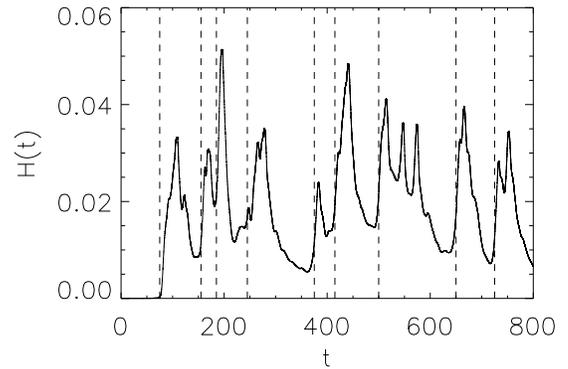}
\caption{The volume integrated heating function, $H(t)$, as a function of time. The dashed vertical lines indicate the start of the different stages, as determined from the $j_z$ contour plots.}
\label{fig:heat}
\end{center}
\end{figure}



\section{Conclusions}
Using a fully 3D MHD simulation, we have demonstrated for
the first time how a local instability in the coronal magnetic field
can trigger large-scale energy release through an avalanche. This involves an initial
instability in one magnetic thread, the nonlinear phase of which leads 
to magnetic reconnection with neighbouring stable threads. The process is repeated until
most of the stable threads have been engulfed, leading to a large, highly structured,
weakly twisted flux rope. Each disruption results in
significant release of magnetic energy and in our simulation approximately three
quarters of the maximum magnetic energy is released in each
thread. This results in a series of heating events, whose height
and width depend on the number of threads disrupted at each stage. The final state shows some aspects that may be ascribed to a turbulent or chaotic system, 
such as a large hierarchy of scales in the final current profiles.

The demonstration here does not yet address many of the aspects present
in the Self Organised Criticality or SOC-type models discussed in the Introduction such as power-law distributions of
energy release as well as the lack of continual driving. In part that is a computational issue (see below), but it is important to 
note the need for avalanches to have varying degrees of efficiency. In the present
simulation, that requires the termination of the avalanche before it can engulf all the
threads. Two aspects are likely to be important. One is the separation between the
threads but this may not be a major issue if all the threads come from discrete photospheric 
magnetic sources. If there is no normal magnetic field between the photospheric sources, then
the field will automatically expand until it touches its neighbouring fields. The second issue
is whether the introduction of threads with opposite twist can \lq block\rq\  the spread of the
avalanche. Preliminary simulations suggest that this can indeed happen, and this will
be presented fully in the future.

While SOC/CA models have been very useful in understanding coronal phenomenon, there are
always concerns about whether their \lq rules\rq\  are consistent with the equations of MHD. A meaningful
examination of avalanches in MHD has been precluded in the past by the large computer 
resources required. Even now, it will be a major challenge to simulate, in proper 3D MHD, the statistical balance between driving and
avalanches that are central to SOC models.

\section*{Acknowledgements}
AWH would like to thank KU Leuven for the kind hospitality where this work was completed. {AWH, PJC and PKB are three of Eric Priest's 
earliest research students and it is a pleasure to acknowledge his contribution to the development of their research careers.}
We acknowledge the financial support of STFC through the Consolidated grants to the University of St Andrews and the University of Manchester. 
This work used the DIRAC 1, UKMHD Consortium machine at the
University of St Andrews and the DiRAC Data Centric system at Durham University,
operated by the Institute for Computational Cosmology on behalf of the
STFC DiRAC HPC Facility (www.dirac.ac.uk). This equipment was funded by a BIS National
E-infrastructure capital grant ST/K00042X/1, STFC capital grant ST/K00087X/1,
DiRAC Operations grant ST/K003267/1 and Durham University. DiRAC is part of the
National E-Infrastructure.

\end{document}